\documentclass[11pt,a4paper,dvips]{scrartcl}
\usepackage{cite,mcite}
\usepackage{graphicx}
\usepackage{subfig}
\usepackage{lcd}
\usepackage{lcd_title,ifthen}
\usepackage{mathptmx}
\usepackage{helvet}
\usepackage{amsmath}
\usepackage{color}
\usepackage{units}
\usepackage{url}
\makeatletter
\def\url@myurlfontstyle{%
  \@ifundefined{selectfont}{\def\UrlFont{\sf}}{\def\UrlFont{\small\ttfamily}}}
\makeatother
\long\def\symbolfootnote[#1]#2{\begingroup%
\def\thefootnote{\fnsymbol{footnote}}\footnote[#1]{#2}\endgroup} 
 \lcdnote{2010-006}
\newlength{\capindent}
\newlength{\capwidth}
\newlength{\figwidth}
\newcommand{\icaption}[2][!*!,!]{\hspace*{\capindent}%
  \begin{minipage}{\capwidth}
    \ifthenelse{\equal{#1}{!*!,!}}%
      {\caption{#2}}%
      {\caption[#1]{#2}}
      \vspace*{3mm}
  \end{minipage}}
\begin{document}
\begin{titlepage}
%
\vskip 35mm
%
\mydocversion
%
\title{A Study of $e^+e^- \rightarrow H^0 A^0 \rightarrow b \bar b b \bar b$ at 3~TeV at CLIC}
%
\author{M. Battaglia\affiliated{1} \affiliated{2}, P. Ferrari\affiliated{3}}
\affiliations{
\affiliation[1]{CERN, Geneva, Switzerland},\\
\affiliation[2]{University of California at Santa Cruz, Santa Cruz, CA, USA}\\
\affiliation[3]{NIKHEF, Amsterdam, The Netherlands}}
%
\date{June 22, 2010}
%
\begin{abstract}
\noindent
The precise determination of the masses of the CP-odd and -even 
heavy Higgs bosons is an important part of the study of Supersymmetry 
and its relation with cosmology through dark matter. This note presents
a determination of the $A^0$ mass with the 
$e^+e^- \to H^0A^0 \to b \bar b b \bar b$ process for a dark matter 
motivated cMSSM scenario with $M_A$ = 1141~GeV at CLIC.
The analysis is performed with full simulation and reconstruction 
at $\sqrt{s}$=3~TeV accounting for beamstrahlung effects.
SM and SUSY backgrounds are considered and the effect 
of the overlay of  $\gamma \gamma \rightarrow {\mathrm{hadrons}}$ 
events on the signal is studied for various assumptions for the 
detector time-stamping capabilities. The di-jet mass resolution is 
improved by applying a kinematic fit. The $A^0$ mass can be 
determined with a statistical accuracy of $\simeq$~3-5~GeV
for 3~ab$^{-1}$ of statistics and 0 to 20~bunch crossings of 
$\gamma \gamma$ background integrated in one event, respectively.
\end{abstract}
%
%
\end{titlepage}
%
%
\section{Introduction}

The CLIC linear collider is expected to probe physics at the multi-TeV scale, 
complementing and extending the reach of the LHC on New Physics beyond the 
Standard Model. If this New Physics is manifested by a rich spectroscopic 
of new particles, as several models predict, their study will be a key part
of the CLIC physics program. In this note we study the pair production 
of heavy supersymmetric Higgs bosons, $e^+e^- \to H^0A^0 \to b \bar b b \bar b$ 
for the parameters of benchmark point~K', defined in the constrained Minimal 
Supersymmetric Standard Model (cMSSM) extension in ref.~\cite{Battaglia:2003ab}. 
This benchmark is characterised by very 
heavy supersymmetric particles, with kinematic thresholds for pair production 
between 2 and 3~TeV. The $H^0$ and $A^0$ bosons have masses of 1141 and 1137~TeV, 
respectively, with a natural width $\Gamma$ = 28~GeV. The accurate determination 
of these masses is essential for outlining the SUSY profile and understanding the 
connection of SUSY with cosmology, through dark matter. The $A^0$ mass plays a 
special role in this respect. In fact, the relation between its mass, $M_A$, 
and that of the lightest neutralino, $M_{\chi^0_1}$, determines the cross section 
of the $\chi^0_1 \chi^0_1 \rightarrow A^0$ annihilation process in the early universe, 
which could have been responsible for determining the amount of relic dark matter
we observe in the universe today~\cite{Ellis:2001msa}. Pair production of heavy 
neutral Higgs bosons at the linear collider has already been studied for several 
scenarios with lighter particles~\cite{uppsala, halcc4}. This analysis is interesting 
for assessing the detector requirements in terms of $b$-tagging efficiency, since 
the signal has four $b$ jets and small production cross section, jet clustering,  
di-jet invariant mass reconstruction and detector time stamping capabilities in 
presence of two-photon machine-induced background.

\section{Event Simulation}

For this study, signal events are generated with {\sc Isasugra~7.69}~\cite{Paige:2003mg} 
and {\sc Pythia~6.125}~\cite{Sjostrand:2006za}. The observed signal cross section at 3~TeV 
is 0.3~fb, accounting for the CLIC luminosity spectrum. The main SM background 
processes are generated with {\sc Pythia~6.125}. In addition, the irreducible inclusive 
$b \bar b b \bar b$ SM background is generated with CompHep~\cite{Comphep} at the 
parton level and subsequently hadronised with {\sc Pythia}. The number of events generated 
for the  various processes considered here and their cross sections are given in 
Table~\ref{tab:mcstat}. 
Beamstrahlung effects are included using {\sc Calypso} for the CLIC~2008 accelerator 
parameters~\cite{Braun:2008zzb}.  We assume the beams to be unpolarised. Events are 
processed through full detector simulation using the {\sc Geant-4}-based 
{\sc Mokka}~\cite{MoradeFreitas:2004sq} program and reconstructed with
{\sc Marlin}-based~\cite{Gaede:2006pj} processors assuming a version of the ILD 
detector~\cite{ild}, modified for physics at CLIC.
A total of $\simeq$100k SM and SUSY events have been generated, fully simulated, 
reconstructed and analysed (see Table~\ref{tab:mcstat}).
\begin{table}
\caption{Summary of simulated samples}
\begin{center}
\begin{tabular}{l|c|c|c}
\hline
Process     & $\sigma$ & Generator & Nb of Events  \\
            & (fb)     &           &  Generated    \\
\hline
$H^0A^0$    &  ~~0.3   & ISASUGRA~7.69+ &  ~3000     \\
            &          & PYTHIA~6.125   &           \\
\hline
Inclusive   & ~32.5    & ISASUGRA~7.69+ &  20000    \\
SUSY        &          & PYTHIA~6.125   &           \\
$H^+H^-$    & ~~0.6    & ISASUGRA~7.69+ &  ~3000     \\
            &          & PYTHIA~6.125   &           \\
$W^+W^-$    & 464.9    & PYTHIA~6.125   &  30000     \\
$Z^0Z^0$    & ~26.9    & PYTHIA~6.125   &  15000     \\
$t \bar t$  & ~19.9    & PYTHIA~6.125   &  15000     \\
$b \bar b b \bar b$& ~~0.4 & CompHEP+   &  ~5000     \\
            &          & PYTHIA~6.125   &            \\ 
$W^+W^-Z^0$ & ~32.8    & PYTHIA~6.125   &  ~7500     \\
$Z^0Z^0Z^0$ & ~~0.4    & PYTHIA~6.125   &  ~2500     \\
\hline
\end{tabular}
\label{tab:mcstat}
\end{center}
\end{table}   
Particle tracks are reconstructed by the combination of a Time Projection 
Chamber and a pixellated Si Vertex Tracker. The momentum resolution is 
$\delta p/p^2$ = 2~$\times$10$^{-5}$~GeV$^{-1}$ and the impact parameter 
resolution $\sigma_{R-\Phi}$ = ($5~\oplus~\frac{15}{p_t {\mathrm{(GeV)}}})$~$\mu$m.
The parton energy is reconstructed using the {\sc Pandora}PFA particle 
flow algorithm~\cite{Thomson:2009rp}. Performances are discussed below.
For the $\gamma \gamma \rightarrow {\mathrm{hadrons}}$ background simulation 
two photon events are generated with the {\sc Guinea~Pig} program~\cite{c:thesis}
using the 2008 nominal CLIC beam parameters at $\sqrt{s}$ = 3~TeV where the hadronic 
background cross section is modelled following Schuler and 
Sj\"ostrand~\cite{c:had0}. The energies of two colliding photons are stored with 
their relevant probability and passed to {\sc Pythia} for the generation of the 
hadronic events. On average, there are 3.3 $\gamma \gamma \rightarrow {\mathrm{hadrons}}$ 
events per bunch crossing (BX) with $M_{\gamma \gamma} >$ 3~GeV.  These are passed through 
the same {\sc Mokka} full detector simulation and overlayed to the particles originating 
from the primary $e^+e^-$ interaction at the reconstruction stage, using the event overlay 
feature in the {\sc Lcio} persistency package~\cite{Gaede:2005zz}. For this analysis, 
the $\gamma \gamma$ background is overlayed only to the signal $H^0A^0$ events, in order to 
study its effect on the signal reconstruction and to the peaking $H^+H^-$ background events.
These samples are reconstructed with different amounts of overlayed background, corresponding 
to different time-stamping performances of the detector. The remaining backgrounds have a flat 
di-jet invariant mass distribution in the signal region and the discriminating variables adopted 
are not significantly affected by the two-photon background.
The comparison of these results allows us to assess some requirements on the detector timing 
capabilities. 

\section{Event Selection and Reconstruction}

At the chosen SUSY parameters, where the $e^+e^- \to H^0A^0$ process is close to its 
kinematic threshold, the production cross section is small. The natural signal-to-background 
ratios to SM processes such as $W^+W^-$ and $t \bar t$ are $\simeq$~1/1500 and 1/60, 
respectively. Therefore, a set of selection criteria with rejection factors of 
${\cal{O}}{\mathrm{(10^3-10^4)}}$ must be applied. There are two main sets of cuts 
which effectively reduce the SUSY and SM backgrounds. 
The first is based on the event shape and jet multiplicity, since two-fermion and gauge 
boson pair production events at 3~TeV are highly boosted, while the heavy Higgs bosons 
of the signal are produced with small kinetic energy, giving 4-jet, spherical events. 
The second is $b$-tagging since all four jets in the signal sample contain $b$ hadrons. 
The $b \bar b b \bar b$ irreducible SM background has a cross section of only 0.4~fb, 
which makes $b$-tagging the single most effective selection criterium to isolate the 
signal.

In order to reject particles which are poorly reconstructed or likely to originate 
from underlying $\gamma \gamma$ events, a set of minimal quality cuts are applied. Only 
particles with $p_t >$~1~GeV are considered, charged particles are also required to have 
at least 12 hits in the tracking detectors and $\delta p/p<$~1. 
The event selection proceeds as follows. First multi-jet hadronic events with little or 
no observed missing energy are selected. We require events to have at least 50~charged 
particles, total reconstructed energy exceeding 2.7~TeV, event thrust below 0.95, 
sphericity larger than 0.04, transverse energy exceeding 1.3~TeV and 
3~$< N_{\mathrm{jets}} <$~6, where $N_{\mathrm{jets}}$ is the natural number of jets 
reconstructed using the Durham clustering algorithm~\cite{Catani:1991hj} with $y_{cut}$ 
= 0.0025. These cuts remove all the SUSY events with missing energy and the 
$e^+e^- \to f \bar f$ events. 

For events fulfilling these criteria, we perform the final jet reconstruction using two 
different approaches and we study the sensitivity of the reconstruction techniques on the 
two-photon background. The first approach is to apply an exclusive 
reconstruction by forcing the event into four jets, as traditionally done in $e^+e^-$ event
reconstruction. Here, we use the Durham clustering algorithm and we shall refer to this as 
the ``4-jet Durham'' method. The second approach uses a more inclusive technique. Since 
$\gamma \gamma$ events give rise to soft jets, it appears advantageous to cluster jets using 
a method which behaves like a cone algorithm and select only the four 
most energetic jets for performing the di-jet reconstruction. We use the anti-kt clustering 
algorithm in cylindrical coordinates~\cite{Cacciari:2008gp}, implemented in the {\tt FastJet} 
package, which we have ported as a processor in the {\tt Marlin} analysis framework. 
The choice of cylindrical coordinates is optimal since the $\gamma \gamma \to {\mathrm{hadrons}}$ 
events are forward boosted, similarly to the underlying events in $pp$ collisions for which 
the anti-kt clustering has been optimised.
For each event, we choose the minimum $R$ value at which the event has exactly four jets with 
energy in excess of 150~GeV. These jets are then used for the reconstruction, while the lower 
energy jets, which are mostly due to the $\gamma\gamma$ events when the background 
is overlayed, are disregarded.  We shall refer to this as the ``semi-inclusive anti-kt'' method.
The effect of $b$-tagging selection is included for the reconstructed jets, requiring four jets 
to be $b$-tagged. For this analysis we use a parametrised response and we assume that the efficiency 
for jets containing decays products of $b$ quarks, $\epsilon_b$, is 0.85 with a mistag probability 
of 0.05 and 0.01 for $c$ and light quarks, respectively. We study the dependence of the result on 
$\epsilon_b$ in Section~4.
\begin{figure}[ht!]
\begin{center}
\includegraphics[width=9.5cm]{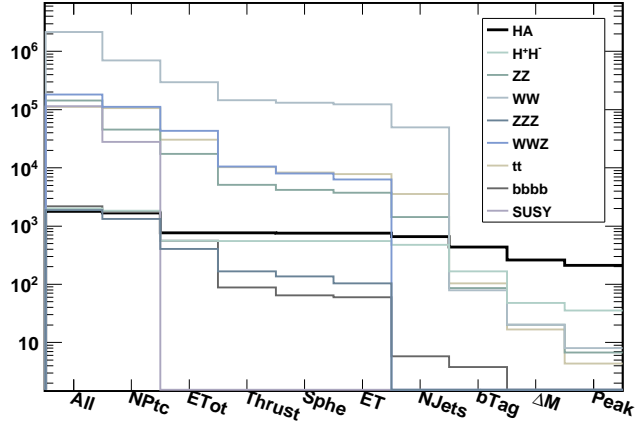}
\end{center}
\caption{Number of selected di-jets for the signal and the main background processes 
at each stage of the event selection procedure}
\label{fig:HACuts}
\end{figure}
Then, the di-jet invariant mass is computed. Since there are three possible permutations
for pairing the four selected jets and the $H^0$ and $A^0$ bosons are expected to be almost 
degenerate in mass, we take the combination minimising the difference $\Delta M$ of the two 
di-jet invariant masses and require $|\Delta M| <$~190~GeV. Since the signal events are 
predominantly produced in the central region while the $\gamma \gamma$ and some of the 
SM background events are forward peaked, we only accept events for which the jet with 
the minimum polar angle, $\theta$, has  $|\cos \theta| <$ 0.92. Finally, events with one 
or more jets having an invariant mass within 12~GeV of the top mass are discarded, to 
reject $t \bar t$ and $H^+ H^- \to t \bar b \bar t b$ backgrounds. The number of selected 
events for the signal and the main background 
processes at each stage of the event selection procedure is shown in Figure~\ref{fig:HACuts}. 
We study the parton energy resolution for jets from signal events fulfilling our selection 
criteria. We compare the jet energy from particle flow to the energy of the $b$ hadron 
which is closest in energy-momentum space. We study the distribution of the quantity 
$\frac{E_{b} - E_{\mathrm{jet}}}{E_b}$ and we determine the r.m.s. of the distribution, 
truncated to include the 90\% of the entries closest to the peak, $\mathrm{RMS_{90}}$. 
We observe a relative parton energy resolution $\mathrm{RMS_{90}}/E_{\mathrm{jet}}$ = 
0.113$\pm$0.002 and 0.112$\pm$0.002  for jets from selected signal $H^0A^0$ events in absence 
of overlayed background using the 4-jet Durham and the semi-inclusive anti-kt algorithms, 
respectively. This resolution is due to both the performance of the particle flow for jet energies 
as high as 1.3~TeV and also the effect of neutrinos produced in semileptonic heavy quark decays 
and escaping the detector.\footnote{Repeating the analysis for $H^0A^0$ events where both bosons are 
forced to decay into light quark and into $b$ quarks without semileptonic decays with escaping 
neutrinos we obtain a relative parton energy resolution $\mathrm{RMS_{90}}$ = 0.088$\pm$0.002 and 
$\mathrm{RMS_{90}}$ = 0.093$\pm$0.004, respectively.} 
The evolution of the $\mathrm{RMS_{90}}/E_{\mathrm{jet}}$ resolution on the $b$-jets with the 
amount of overlayed $\gamma \gamma$ events for the two algorithms is summarised in 
Table~\ref{tab:sigmaJ}.
\begin{table}
\caption{$\mathrm{RMS_{90}}$ relative energy resolution for $b$-jets in signal events 
reconstructed with the 4-jet Durham clustering and the semi-inclusive anti-kt algorithm 
for different amounts of overlayed $\gamma \gamma$ background.} 
\begin{center}
\begin{tabular}{l|c c}
Nb of BX & \multicolumn{2}{c}{$\mathrm{RMS_{90}}/E_{\mathrm{jet}}$} \\
of overlayed $\gamma \gamma$ & 4-jet & semi-inclusive \\
                             & Durham & anti-kt \\
\hline
~0       & 0.113$\pm$0.002 & 0.112$\pm$0.002 \\
20       & 0.149$\pm$0.003 & 0.129$\pm$0.002 \\
40       & 0.170$\pm$0.003 & 0.133$\pm$0.002 \\
60       & 0.183$\pm$0.003 & 0.140$\pm$0.002 \\
\hline
\end{tabular}
\label{tab:sigmaJ}
\end{center}
\end{table}

The Higgs mass resolution is studied by fitting the di-jet invariant mass distribution with 
the sum of two Breit-Wigner functions, describing the $A^0$ and $H^0$ signals, folded with 
a Gaussian resolution term. In this fit we set three parameters free: the $A^0$ mass, $M_A$, 
the Gaussian resolution term, $\sigma_M$, and the number of signal events, $N_{\mathrm{signal}}$, 
while keeping the $H^0$ to $A^0$ mass splitting and the natural widths of the bosons fixed to 
their values in the SUSY model. We obtain a di-jet invariant mass resolution $\sigma_M$ = 
(68.5$\pm$6.2)~GeV with the 4-jet clustering and (76.3$\pm$6.3)~GeV using the semi-inclusive 
anti-kt.

\begin{figure}
\begin{center}
\includegraphics[width=7.5cm]{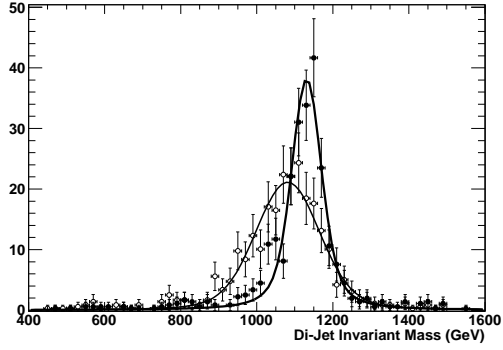}\\
\end{center}
\caption{Di-jet invariant mass distribution  for signal events before (open circles) and after 
(filled circles) kinematic fit using the semi-inclusive anti-kt jet clustering.}
\label{fig:HAMass}
\end{figure}

In order to improve the di-jet mass resolution, we apply a constrained kinematic fit.
We use the port of the {\sc Pufitc} kinematic fit algorithm~\cite{pufitc} to the 
{\sc Marlin} framework. {\sc Pufitc} was originally developed for $W^+W^-$ reconstruction 
in DELPHI at LEP-2 and it has been successfully applied for the reconstruction of simulated 
linear collider events at lower energies~\cite{halcc4}. The kinematic fit adjusts the momenta 
of the four jets as $p_F = a p_M + b p_B + c p_C$, where $p_M$ is the jet momentum from 
particle flow, $p_B$ and $p_C$ are unit vectors transverse to $p_M$ and to each other and 
$a$, $b$ and $c$ free parameters. The rescaled jet momenta minimise the quantity
$\sum_i (\frac{(a_i-a_0)^2}{\sigma_a^2} + \frac{b_i^2}{\sigma_b^2} + \frac{c_i^2}{\sigma_c^2})$,
where $a^0$ is the expected energy loss parameter, $\sigma_a$ the energy spread and 
$\sigma_b$, $\sigma_c$ the transverse energy spreads as in the notation used above.
For this analysis, we impose the constraints $p_x$ = $p_y$ = 0 and $E \pm |p_z| = \sqrt{s}$, 
where the last conditions accounts for beamstrahlung photons radiated along the beam 
axis. We take the values of the energy spread from simulation. We only accept events with a 
kinematic fit $\chi^2 <$~10. After the kinematic fit, the 
relative jet energy resolution $\mathrm{RMS_{90}}/E_{\mathrm{jet}}$ improves to 
0.094$\pm$0.002 and to 0.103$\pm$0.002 without and with 20~BX of $\gamma \gamma$ background, 
respectively. The di-jet invariant mass resolution improves by more than a factor of two to 
$\sigma_M$ = (27.7$\pm$~4.8)~GeV and (29.8$\pm$4.7)~GeV using the 4-jet and the semi-inclusive 
anti-kt method, respectively (see Figure~\ref{fig:HAMass}).
\begin{table}
\caption{Higgs mass experimental width after kinematic fitting for different 
amounts of overlayed $\gamma \gamma$ background using the 4-jet Durham clustering
and the semi-inclusive anti-kt algorithm.} 
\begin{center}
\begin{tabular}{l|cc|}
Nb of BX & \multicolumn{2}{|c|}{Kinematic Fit} \\
of overlayed $\gamma \gamma$ &  \multicolumn{2}{|c|}{$\sigma_{M_{jj}}$ (GeV)} \\
                             & 4-jet & semi-incl. \\
\hline
~0       & ~27.7$\pm$~4.8 &   29.8$\pm$4.7\\
~5       & ~32.0$\pm$~5.0 &   30.1$\pm$5.0\\
20       & ~54.0$\pm$~8.3 &   34.5$\pm$6.7\\
40       & ~72.2$\pm$~7.4 &   45.4$\pm$5.6\\
60       & ~78.6$\pm$10.9 & 52.5$\pm$8.2 \\
\hline
\end{tabular}
\label{tab:sigmaM}
\end{center}
\end{table}
The use of a kinematic fit also mitigates the effect of the overlayed $\gamma \gamma$ 
events on the di-jet mass resolution. Since we do impose the nominal centre-of-mass 
energy, allowing for beamstrahlung, jet energies are rescaled in the fit. 
The Gaussian resolutions of the di-jet invariant masses obtained for the raw jet 
energy and momentum from the particle flow and that for the rescaled values after 
the kinematic fit are given in Table~\ref{tab:sigmaM} for the two jet algorithms 
and different amounts of overlayed backgrounds. Up to about 20-30~BXs the kinematic 
fit ensures a good di-jet invariant mass resolution. Above 40~BXs of background, 
corresponding to about 1.75~TeV of total energy from 
$\gamma \gamma \rightarrow {\mathrm{hadrons}}$ events overlayed to 
the $e^+e^- \to H^0A^0$ event, the di-jet mass resolution quickly degrades for the 
4-jet cluster while it remains reasonably good for the  semi-inclusive anti-kt 
algorithm.

\section{Results}

We consider the results of the analysis of a data set corresponding to an 
integrated luminosity of 3~ab$^{-1}$ taken at $\sqrt{s}$ = 3~TeV, corresponding 
to the production of $\simeq$~800 $e^+e^- \to H^0A^0$ events.
\begin{figure}
\begin{center}
\begin{tabular}{cc}
\subfloat[0 BX]{\label{fig:00BX}\includegraphics[width=7.0cm]{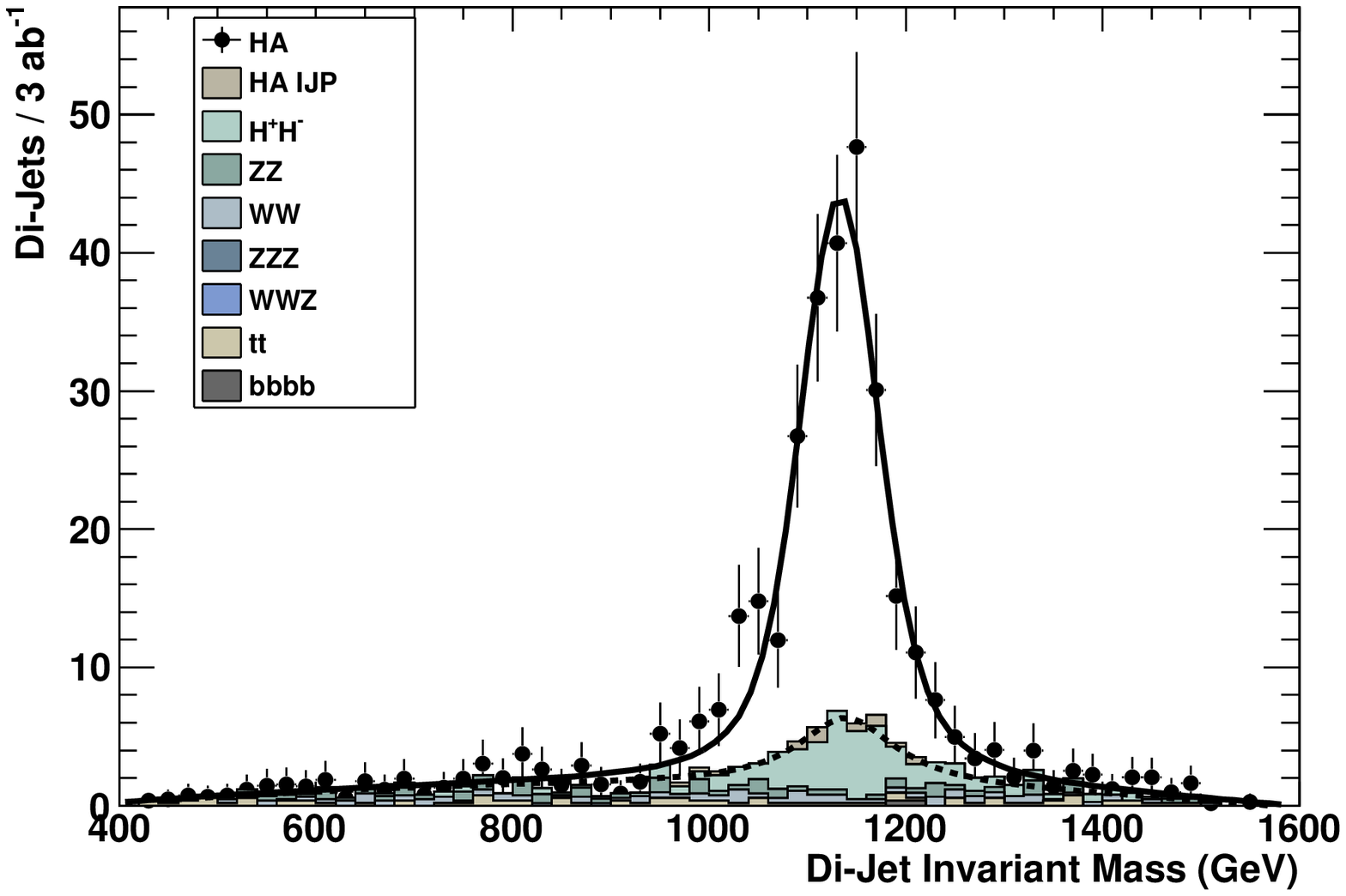}} &
\subfloat[20 BX]{\label{fig:20BX}\includegraphics[width=7.0cm]{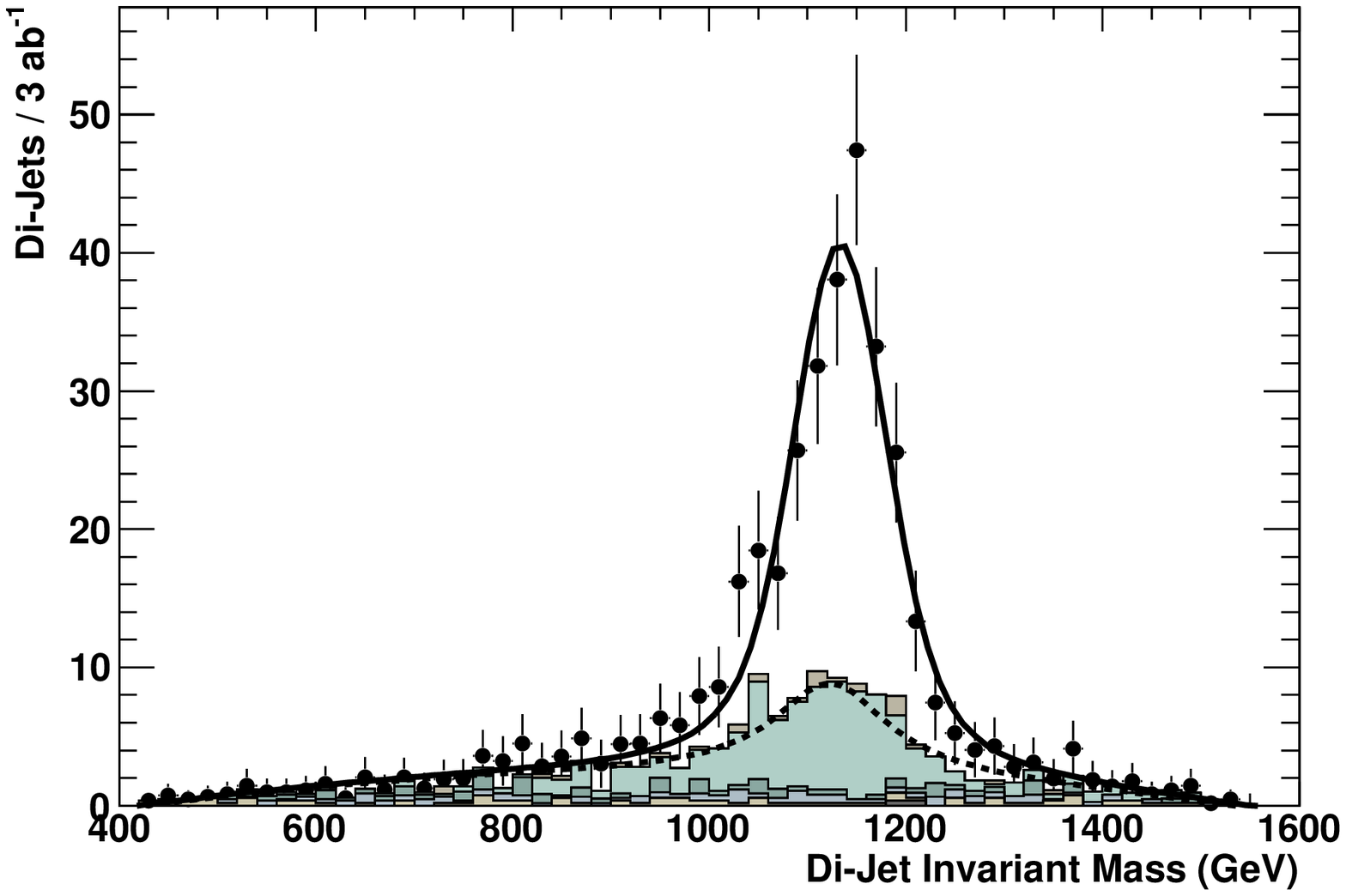}}\\
\subfloat[40 BX]{\label{fig:40BX}\includegraphics[width=7.0cm]{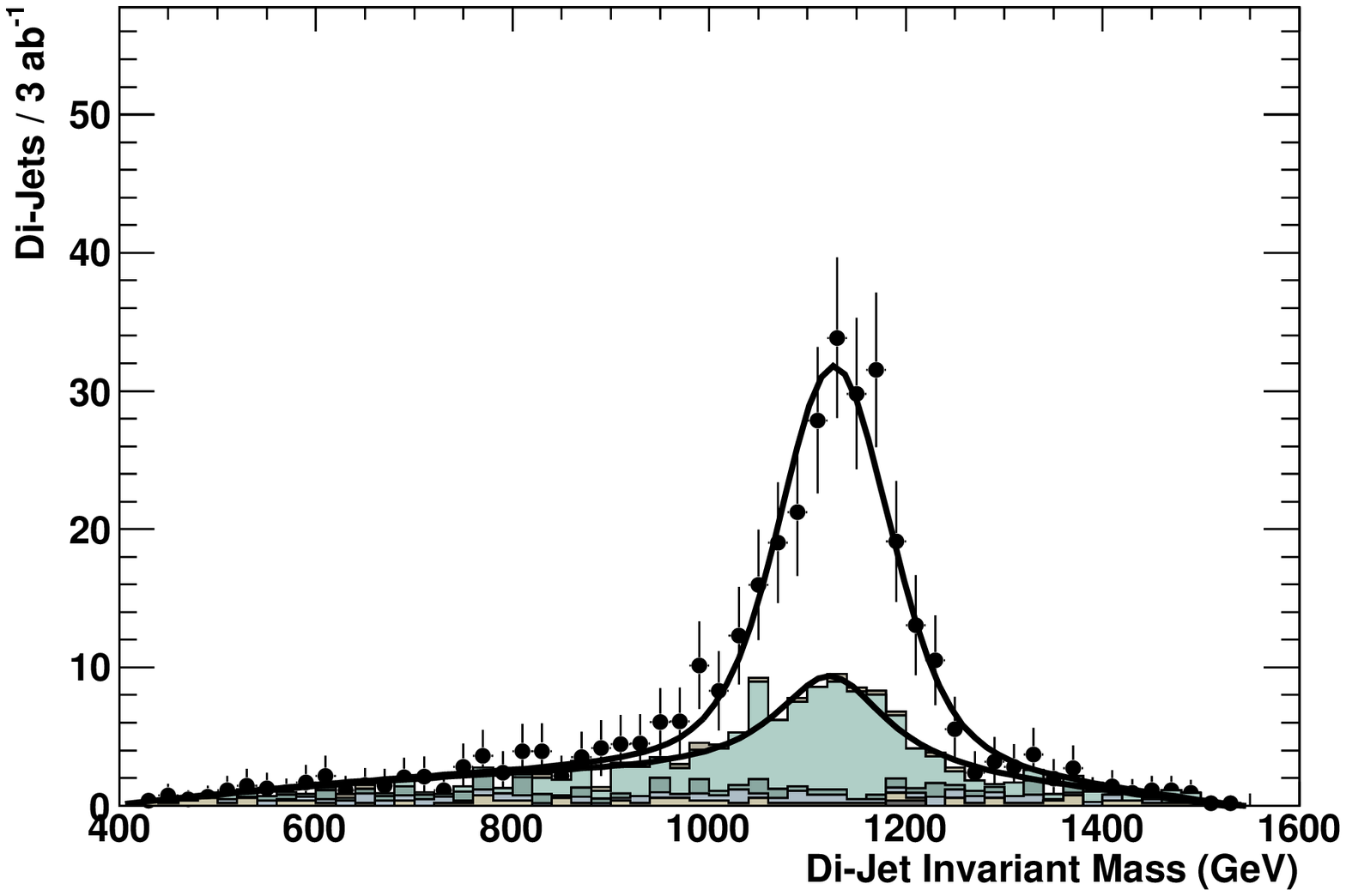}} &
\subfloat[60 BX]{\label{fig:60BX}\includegraphics[width=7.0cm]{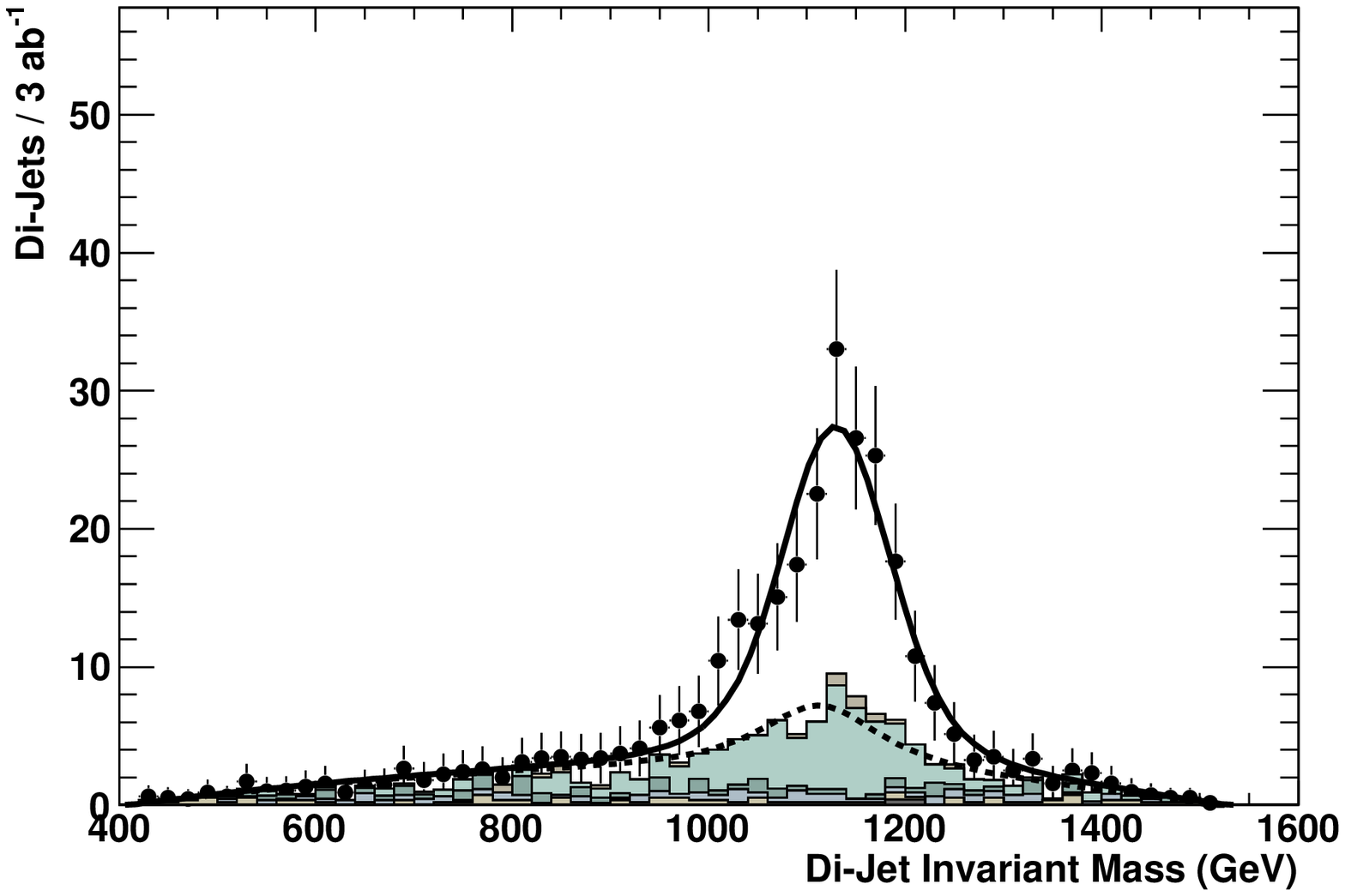}}\\
\end{tabular}
\end{center}
\caption{Di-jet invariant mass distribution for (a) no, (b) 20, (c) 40 and 
(d) 60 BXs of overlayed $\gamma \gamma$ background for 3~ab$^{-1}$ of integrated 
luminosity at $\sqrt{s}$=3~TeV. The $HA$ signal, the $HA$ events with incorrect jet 
pairing and the various background components are highlighted.}
\label{fig:HAMassBX}
\end{figure}
The invariant mass distribution of the two di-jets in events selected by the criteria 
discussed above is shown in Figure~\ref{fig:HAMassBX} for various amounts of 
$\gamma \gamma$ background overlayed. The SM background is  small and flat in the signal 
region. The only peaking background is represented by 
$H^+ H^- \to t \bar b \bar t b$ events fulfilling the selection criteria. These poorly 
reconstructed charged Higgs decays passing the anti-top and jet selection criteria
give di-jet masses centred at approximately the same mass as the $H^0$ and $A^0$ bosons, 
since $M^{H^{\pm}} =$ 1148~GeV in this model, but with a broader distribution. We repeat 
the fit to the di-jet mass 
distribution adding a smooth background, parametrised by a second order polynomial and 
a peaking background term, parametrised by a Breit-Wigner distribution. We obtain the 
parameters of these distributions by fitting the distributions for the corresponding 
samples. The fitted number 
of signal, $N_{\mathrm{signal}}$, and background, $N_{\mathrm{bkg}}$, events and the 
$M_A$ mass value after the kinematic fit for the various $\gamma \gamma$ overlay conditions 
are summarised in Table~\ref{tab:fitM}. We observe a significant advantage in using the 
semi-inclusive anti-kt method over the more traditional 4-jet clustering.
\begin{table}
\caption{Number of  selected background and signal di-jets and value of the $A^0$ mass 
after kinematic fitting for different amounts of overlayed $\gamma \gamma$ background
for 3~ab$^{-1}$ of integrated luminosity at $\sqrt{s}$=3~TeV.} 
\begin{center}
\begin{tabular}{l|c|c|c c}
Nb.\ of BXs & $N_{\mathrm{bkg}}$ & $N_{\mathrm{signal}}$ &  \multicolumn{2}{|c|}{$M_A$ (GeV)} \\
of overlayed $\gamma \gamma$ &  &  & 4-jet & semi-incl. \\
\hline
~0       & ~76 & 222$\pm$19 & 1137.4$\pm$3.3 & 1136.7$\pm$3.4\\
~5       & ~77 & 224$\pm$20 & 1144.4$\pm$3.8 & 1135.9$\pm$3.7\\
20       & 102 & 208$\pm$20 & 1160.7$\pm$6.9 & 1139.9$\pm$5.4\\
40       & ~96 & 183$\pm$20 & 1167.2$\pm$8.2 & 1134.1$\pm$7.2\\
60       & ~97 & 167$\pm$21 & 1170.6$\pm$9.7 & 1132.9$\pm$8.6\\
\hline
\end{tabular}
\label{tab:fitM}
\end{center}
\end{table}
While performances are comparable in absence of background,
the statistical accuracy of the mass measurement is more stable 
with the amount of overlayed background when using the semi-inclusive 
anti-kt method. The $A^0$ mass is measured with a statistical accuracy of 
$\simeq$~3~GeV in absence of $\gamma \gamma$ background, which increases to 
$\simeq$~5~GeV for 20~BXs of overlayed background. 
A $\simeq$~3~$\sigma$ systematic upward shift of the fitted mass value for 
20 or more bunch crossings of background, which appears for the 4-jet clustering 
method, is removed by using the semi-inclusive anti-kt clustering and the fitted 
mass remains within less than 1~$\sigma$ from its simulated value. 
\begin{figure}[ht!]
\begin{center}
\includegraphics[width=7.5cm]{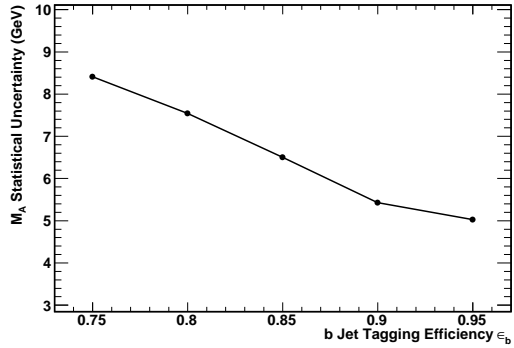}\\
\end{center}
\caption{Statistical uncertainty on $M_A$ as a function of the $b$-tagging 
efficiency per jet, $\epsilon_b$, for 20~BX of $\gamma \gamma$ background 
overlayed.}
\label{fig:effb}
\end{figure}
Finally, we repeat the analysis by changing the $b$-tagging efficiency 
per jet, $\epsilon_b$, in the range 0.75 $< \epsilon_b <$ 0.95 and study the 
change in the statistical accuracy on $M_A$ as a function $\epsilon_b$ 
(see Figure~\ref{fig:effb}).

\section{Conclusions}

The process $e^+e^- \to H^0A^0 \to b \bar b b \bar b$ has been studied
for the parameters of a dark matter motivated, high mass SUSY scenario
with $M_A$ = 1141~GeV. The analysis is based on fully simulated and 
reconstructed events and includes both physics backgrounds and 
machine-induced $\gamma \gamma$ events. CLIC at 3~TeV can 
study the production of pairs of heavy Higgs particles with high 
accuracy up to masses close to the production kinematic threshold. 
Event shape discriminating variables and 
jet flavour tagging efficiently reduce the SM and SUSY backgrounds 
to achieve a good signal-to-background ratio in the signal region at 
high mass. A semi-inclusive jet clustering using a cone-like algorithm 
in cylindrical coordinates helps in reducing the impact of overlayed events 
and ensure a stable, unbiased mass determination up to largest number of 
overlayed bunch crossings considered in this study. 
Due to the favourable kinematics of the signal events, the di-jet invariant 
mass resolution can be significantly improved by performing 
a constrained kinematic fit, accounting for beamstrahlung.
This fit further mitigates the impact of overlayed $\gamma \gamma$ 
background up to $\simeq$~20-30~BXs, corresponding to detector 
time stamping accuracies of 10-15~ns. Under these conditions, the 
$A^0$ mass can be determined with a statistical accuracy of $\pm$3-5~GeV
for an integrated luminosity of 3~ab$^{-1}$. 
This accuracy matches that required for obtaining predictions on the relic 
dark matter density in the universe from collider data with a precision 
comparable to that presently achieved in the study of the cosmic microwave 
background with the WMAP satellite data~\cite{Larson:2010gs}.

The results reported in this note can be further improved. The particle flow 
algorithm still needs to be optimised for jets of $\cal{O}$(1~TeV). 
Explicit $b$-tagging has yet to be performed and can profit from the 
considerable decay length of the energetic $b$ hadrons. The jet clustering 
algorithm can be further optimised for the reconstruction of events in presence 
of a diffuse background, as in the case of a large number of overlayed 
$\gamma \gamma$ events. 

%
%
%
\section{Acknowledgements}

We are grateful to the colleagues who contributed to this study. In particular,  
D.~Schulte provided the luminosity spectrum and the $\gamma \gamma$ background 
events. E.~Boos and S. Bunichev helped with the generation of events with 
CompHep. J.~Quevillon studied jet clustering of $HA$ events at the early stage
of the analysis, while M.~Cacciari guided us in the use of the anti-kt algorithm  
implemented in FastJet. D.~Schlatter and A.~Nomerotski carefully read the 
manuscript.
%
%

\end{document}